# Coupling between switching regulation and torque generation in bacterial flagellar motor

Fan Bai[1], Tohru Minamino[1], Zhanghan Wu[2], Keiichi Namba[1*], Jianhua Xing[2*]

1.Graduate School of Frontier Biosciences, Osaka University, 1-3 Yamadaoka, Suita, Osaka 565-0871, Japan

2.Department of Biological Sciences, Virginia Tech, Blacksburg, Virginia, 24061-0406, USA

*To whom correspondence should be addressed.

E-mail: jxing@vt.edu, keiichi@fbs.osaka-u.ac.jp

ABSTRACT

The bacterial flagellar motor plays a crucial role in both bacterial locomotion and chemotaxis. Recent experiments reveal that the switching dynamics of the motor depends on the motor rotation speed, and thus the motor torque, non-monotonically. Here we present a unified mathematical model which models motor torque generation based on experimental torque-speed curves and torque-dependent switching based on the conformational spread model. The model successfully reproduces the observed switching rate as a function of the rotation speed, and provides a generic physical explanation independent of most details. A stator affects the switching dynamics through two mechanisms: accelerating the conformation flipping rates of individual rotor switching units, which favours slower motor speed and thus increasing torque; and affecting more switching units within unit time, which favours faster speed. Consequently, the switching rate shows a maximum at intermediate speed. Our model predicts that a motor switches more often with more stators. The load-switching relation may serve as a mechanism for sensing the physical environment, similar to the chemotaxis system for sensing the chemical environment. It may also coordinate the switch dynamics of motors within a cell.

Author contributions: F.B., T.M, K.N. and J.X. designed research; F.B., Z. W. and J.X. performed research, analyzed data; and wrote the paper.

## INTRODUCTION

Most bacteria swim by rotating their flagella. Using the transmembrane electrochemical proton (or sodium) motive force to power the rotation of the bacterial flagellar motor (BFM), free-swimming bacteria can propel their cell body at a speed of 15-100 μm/s, or up to 100 cell body lengths per second (1, 2). Box 1*A* gives a schematic illustration of the key components of the *E. coli* BFM derived from collected research of electron microscopy, sequencing and mutational studies. In the extracellular part of the cell, a long flagellar filament (about 5 or 10 times the length of the cell body) is connected to the motor through a universal joint called the hook.

Under the hook is the basal body, which functions as the rotor of the BFM, and spans across the outer membrane, peptidoglycan and inner membrane into the cytoplasm of the cell (reviewed in (1, 2)). A circular array of 8-11 stator complexes locates around the periphery of the rotor. Each complex functions independently as a torque generation unit. Protons flow from the periplasm to the cytoplasm through a proton channel of the stator complex, which is thought to drive a conformational change of the cytoplasmic domain of the stator complex, interacting with the C-terminal domain of one of the 26 copies of FliG monomers on the rotor to generate torque.

Like macroscopic machines, the torque-speed relationship of a BFM well characterizes its power output under external loads, and indicates the energy conversion efficiency. For *E. coli*, the motor torque remains approximately constant up to a 'knee' velocity of 170 Hz, then drops abruptly to zero at about 300 Hz (3). The sodium-driven flagellar motor shows a similar torque-speed relation with a higher 'knee' speed and zero-load speed (4). The BFM torque-speed curve has also been measured with increasing stator numbers (5), revealing a high duty ratio when individual torque-generating units cooperating with each other.

An *E. coli* BFM can stochastically switch its rotation between clockwise (CW) and counter-clockwise (CCW) directions. When most of the motors on the membrane spin CCW, flagellar filaments form a bundle and propel the cell steadily forward. When a few motors (can be as few as one) spin CW, flagellar filaments fly apart and the cell tumbles. Therefore the cell repeats a 'run'-'tumble'-'run' pattern to perform a biased random walk. Regulation of the motor switching is part of the well-studied chemotaxis system. The latter has long served as a prototype for understanding signal transduction pathways. External signals such as chemical attractants and repellents, temperature, and pH regulate the concentration of the phosphorylated form of a signalling protein CheY, with CheY-P binding to the rotor biasing to CW rotation. Recent studies show that a secondary messenger molecule di-c-GMP also modulates the switching dynamics. Berg and coworkers further show that the motor can response to mechanical signals as well (6, 7). A BFM switches firstly more frequently then less upon decreasing the mechanical load on the motor, with the maximum around rotation speed 100-150 Hz. It is intriguing how the motor integrates chemical and mechanical signals and responses accordingly.

Various mathematical models have been proposed to explain the torque generation and switching mechanisms of the BFM (8-12). With only properties generic to typical protein motors, Xing et al. (8) used a simple model to successfully explain the observed BFM torque-speed curves. As emphasized in (8), the essential physics revealed by the observed torque-speed relations is the interplay between two time scales, which are affected by both the external load and ion motive forces. For the switching mechanism, Duke et al. (13) proposed a conformational spread model for the motor switching dynamics. The model emphasizes that the rotor, as a whole, does not switch between CW and CCW states instantaneously. Instead, each rotor switching unit (RSU) has to switch individually and cooperatively at a finite speed. Recent experimental studies observed the intermediate rotor conformations predicted by the model (10). The conformational spread model successfully explains the ultra sensitivity, transient state, switching interval distribution and finite switching time of the switch complex (10, 13). All these existing model studies treat the switching and torque generation processes as two separate problems.

In this work, we present a unified torque generation-switching model to reveal how coupling between the two processes explains the observation of Berg and coworkers. We will further discuss possible physiological functions of the coupling.

## RESULTS

### Model:

To describe the coupling between torque generation and motor switching, we adopt and generalize the conformational spread model proposed by Bray and Duke in the following aspects:

**1)** Each RSU can be in one of four states. The switch complex is considered to be a ring of 26 identical RSUs. Each RSU possesses a single binding site to which a CheY-P molecule can be bound (*B*) or not bound (*b*), and has two conformations, active (*A*, CW rotation) or inactive (*a*, CCW rotation) (see Box 1B). Box 1C presents a free-energy diagram of the four states of the RSU and the transitions between them. To reproduce the ultra sensitivity of the motor switch, interactions between adjacent RSUs favour pairs with the same conformation. The model assumes that the free energy of the interaction is lowered by $E_j$ for any like pair compared to any unlike pair, independent of CheY-P binding. These interactions add 0, $2E_j$ or $-2E_j$ to the free energy of a conformational change, depending on the state of adjacent RSUs. Above a critical value of $E_j$ the ring spends most of time in a coherent state (completely CCW or completely CW)(see reference (10, 13) for detailed discussions) with occasionally stochastic switching between CCW and CW configurations. Switches typically occur by a single nucleation of a new domain, followed by conformational spread of the domain, which follows a biased random walk until it either encompasses the entire ring or collapses back to the previous coherent state.

2) The total torque on the rotor is a sum of the torques exerted by individual stators obtained from the experimental torque/speed relations (see Box 1D).

3) Switching rates between the two conformational states are affected by the conformation of the neighboring RSUs (see Box 1C), as in the conformational spreading model. Additionally, the instant torque a stator imposing on the directly interacting RSU also accelerates the switching rates of the latter (see Box 1E). Physically the rotor-stator interaction, especially the electrostatic interactions may stabilize the transition state of the switching process.

**Simulated motor switching traces and speed records**

Figure 1*A* shows a typical 30 seconds switching trace of time dependence of the rotor angle $\theta(t)$ from our model. The model successfully captures the stochastic nature of motor's step advancement and switching dynamics. This trace is generated with vanishing external load. Figure 1*B* gives the speed record of the same trace. Motor angular positions are converted to instantaneous speed by dividing the difference between successive angles by the sampling time, 1/3000 s. To reduce noise, the record of speed *vs.* time is 40 points moving average filtered before further analysis (same as what used experimentally (7)). Consistent with our previous studies (10), there are frequent complete switching between CW and CCW states and incomplete switching to intermediate speed levels. Switching event between CW and CCW states takes place non-instantaneously, but with a measureable finite switching time.

## Switching under load

Figure 1 *C-D* shows that the model successfully reproduces the observed load-switching dependence, especially the non-monotonic feature: the switching rate first increases with motor speed, after reaching a maximum value, starts to decrease when speed further increases. Here the switching rate constants are defined as inverse of the average dwelling time in a state. Qualitatively our model provides a generic physical explanation of the non-monotonic feature of the load-switching relationship. A stator affects the switching dynamics through two mechanisms: accelerating the conformation flipping rates of individual RSUs, which favours slower speed and thus increasing torque; affecting more RSUs within unit time, which favours faster speed. Consequently, the load-switching relation shows a maximum at intermediate speed. This mechanism is robust against most details of the model such as parameter values, with the only requirements that the rotor switching takes finite time, and the motor torque is a non-increasing function of the rotation speed in both directions. The generality of our results was verified by reproducing the experimental load-switching relationship equally well using a different form of CW state torque-speed relationship in the model (symmetrical as the CCW state).

To be more quantitative, the dynamics of our model can be characterized by two time scales: $t_{step} = \Delta\theta / v$, which specifies how fast the motor moves, and is proportional to the average RSU number a given stator interacts within a unit time, and $t_{flip} = 1/W_{flip}(\tau) \propto \exp(-|\tau| \times \delta)$, which specifies how fast a single FliG flips with torque assistance, can be defined as the reciprocal of the FliG flipping rate. Figure 2*A* shows the $t_{step}$, $t_{flip}$ dependence on motor speed $v$ with the parameter set used to reproduce the load-switching relationship. To simplify the calculation of $t_{flip}$, here for t we use $t_0^a$ the steady torque contributed by each single stator when the ring is in a coherent CCW state.

The external load affects the two time scales $t_{step}$ and $t_{flip}$ strongly and oppositely. When the external load decreases, the motor speed increases ( $t_{step}$ decreases) and the torque decreases ( $t_{flip}$ increases). Consider two extreme cases. When the motor works with extremely high load, $t_{step} >> t_{flip}$, torque-assisted flipping (for both directions) of the RSU takes place a lot, but the motor seldom steps forward or backward. On the other hand, when motor works with extremely low load, $t_{step} << t_{flip}$, motor stepping events takes place much more often than flipping of individual RSUs. Figure 2B shows that the global switching rate is maximized while the motor works with medium external load, where torque-assisted flipping and speed-enhanced interactions are combined. Figure 2*C* shows the averaged flipping dynamics of 10 randomly selected RSUs on the rotor ring. As expected, the maximum global flipping rate is not achieved with maximum external load, but with medium load, where high torque and fast rotation are both present. Figure 2*D* shows snapshots of the ring activity at high sampling frequency (50000 Hz). Again, the ring is the most active at the medium external load, but not at the largest external load.

**Model predicts that motor switches more often with more stators**

As in our model, motor switching senses the external load through instant torque generated by each stator complex. The more stator units the system has, the higher global flipping rate the ring switches with. Our model predicts that motor switches more often when the system has more stators around the rotor (Figure S1). The dependence is almost linear at both high and low external loads. This prediction awaits experimental tests.

## SUMMARY AND DISCUSSIONS

We have presented a simple mathematical model to explain the load-switching relationship of the bacterial flagellar motor. The conformational spread model, which has successfully reproduced the switching dynamics of the BFM, has been extended to include torque generation and motor movement. The key ingredients of our model are that the instant torque applied from a single stator to each RSU changes its flipping rates, and it takes finite time for the rotor to switch. Van Albada et al proposed that a conformational change of the long helical filament contributes to the coupling (14). Unfortunately this elegant model is inconsistent with the conformational spread model, and the fact that *E. coli* flagellar motor without the flagellar filament has been used for measurements of both motor speed and torque.

The significance of the discovery of Berg and coworkers is that bacteria like *E. coli* can regulate cell motility based on mechanical signaling besides other well studied mechanisms, e.g., chemotaxis. It remains to examine the physiological implications of this observation. Here we provide some possibilities.

Bacteria like *E. coli* and *Salmonella* are propelled by several helical flagellar filaments, each driven at its base by a BFM. A tumble and reorientation of the cell can be caused by as few as one motor rotating CW direction. However, not all situations with one motor rotating CW lead to a tumble of the cell body. For most of time, tumble needs more motors rotating CW. The load-switching relationship may ensure that stochastic switching of one of the motors, which is unlikely directly stimulated by environmental changes, does not cause too frequent tumbling of the cell. A flagellum rotating opposite to others in the bundle experiences uprising external load from the latters. The uprising load helps this motor switch back to CCW state and then the external hindrance vanishes. Only when obvious environmental changes occur, CheY-P concentration rises in the cytoplasm of the cell. This enhances the CW bias of all of the motors and causes rational tumbling of the cell. Therefore we suggest the load-switching relationship acts as a noise suppressor, assumes majority notes, filtering out false tumbling signal due to stochastic switching of one of the multiple motors on the cell surface. The load-dependent cooperative switch provides a cooperative mechanism to explain why motors in a cell switch almost at the same time, a puzzle raised by Turner et al. (15).

Cluzel et al. (16) demonstrated that the motor responses to CheY-P concentration ultrasensitively. An ultrasensitive system might switch back and forth undesirably with the signal (CheY-P concentration) fluctuating around the transition value. The load-induced cooperativity among motors, together with the finite time dynamics of the single motor switch dynamics, filter out local high frequency transient CheY-P fluctuations, and thus increase the motor sensitivity to CheY-P concentration changes.

Within crowded environment (upon cluster forming or under spatial constraint) (17, 18), bacteria may sense the surroundings through the mechanical load, and adjust their moving pattern accordingly. The load-dependent switch may explain the observed variation of the tumbling frequency within a cluster of bacteria (19).

MATERIAL AND METHODS

Below are some model details.

Transition rate constants for conformational change are expressed as:

$k_{a\to A} = \omega_{flip}(\tau)\times \exp(0.5\Delta G(a\to A)/k_B T)$ ,

$k_{A\to a} = \omega_{flip}(\tau)\times \exp(-0.5\Delta G(a\to A)/k_B T)$ , where ΔG is the overall free energy difference of the RSU under consideration between the *A* and *a* conformations, and holds one of six values: $\pm E_a$, $\pm(E_a + 2E_j)$ or $\pm(E_a - 2E_j)$, $k_B T$ is the Boltzmann's constant multiplying temperature. As discussed in previous works (10, 13), to achieve good sensitivity with a rapid kinetics, the optimal performance of the switch complex is obtained when $E_a \approx 1\ k_B T$ and $E_j \approx 1\ k_B T \ln M$ (*M* is the size of the ring = 26 RSUs on the ring).

We further assume that the RSUs in direct contact with stators have a torque-dependent switching parameter. Here we use the same formula as Yuan and Berg (7), $\omega_{flip}(\tau) = \omega_0 \times \exp(|\tau| \times \delta)$ for RSUs interacting with a stator (see also Figure 1*E*), and $\omega_{flip}(\tau) = \omega_0$ for RSUs which are free, where $\delta$ is a scaling factor specifying the strength of the torque dependence. Unlike the work of Yuan and Berg, here $\tau$ is the instant torque an individual stator applies, and it only affects the RSU it contacts. These are essential for our model.

The free energy associated with CheY-P binding depends only on the conformation of the binding RSU, but not on adjacent RSUs. The binding rates are expressed as: $k_{b\to B} = k_{ligand} c / c_{0.5}$ , $k_{B\to b} = k_{ligand} \exp(-\Delta G^*(b\to B)/k_B T)$ ,

where c is the concentration of cytoplasmic CheY-P, $c_{0.5}$ is the concentration of CheY-P required for neutral bias, and $\Delta G^{*}(b \rightarrow B)$ is the free energy associated with binding when CheY-P is at the concentration $c_{0.5}$. We choose $k_{ligand} = 10\ s^{-1}$ based on the experimentally determined CheY-P binding rate, consistent with previous modeling of the switch complex (13). The free energy of CheY-P binding is $\Delta G(b \rightarrow B) = \Delta G^{*}(b \rightarrow B) - \ln(c / c_{0.5})$ for an arbitrary CheY-P concentration.

**Calculation of instant torque generated by individual stators:**

A detailed model such as our original one using continuum-Fokker-Planck approach can give the instant torque generated by each individual stator (8, 9). However, existing structural information is not sufficient to specify the exact shape of the stator-rotor interaction potential required in the continuum-Fokker-Planck method. Furthermore, to explain the observed load-dependent switching dynamics alone, it is both computationally expensive and unnecessary to specify the torque generation details a continuum-Fokker-Planck model requires. Therefore we adopt a strategy that uses experimentally determined results as much as possible. Berg and coworkers have measured the torque-speed relations for *E. coli* BFM in the CW and CCW directions under steady rotation conditions (3, 20). For later discussion we denote the measured two torque-speed relations in the CW and CCW directions $\tau^{A}(v)$ and $\tau^{a}(v)$, respectively. In these cases one can assume that the rotor conformation is in one of the two coherent states. However, during the switch process the rotor accesses a larger conformational space transiently. There are $2^{26} \sim 6.7\times10^{7}$ possible rotor configurations. The number is large even taking into account some degeneracy due to symmetry. Here we introduce a mean field approximation to allow specifying the torque-speed relations for all these configurations from the only two measured curves. We assume that each stator functions independently, and contributes to the overall motor torque additively. For a given stator, the remaining stators function only as additional effective external torques. For a motor with $N$ stators at a given time with a rotation speed $v$, the force balance relation gives, $\tau_{motor}(v) = \sum_{\alpha=1}^{N} \tau_\alpha(v) = \zeta v$, where $\tau_\alpha(v)$ is the instant torque of stator $\alpha$, and $\zeta$ is the drag coefficient of the external load. A corollary of the approximation is a scaling relation of the steady state motor torque-speed curves ( $\tau^A(v)$ and $\tau^a(v)$ ) for different $N$, $\tau_{motor}(v, N_1)/N_1 = \tau_{motor}(v, N_2)/N_2$. Figure S2 show that the data points of Ryu et al. (5) indeed collapse well to a single curve, which supports the validity of the mean field approximation. One can define this normalized curve as the standard torque-speed relationship of a single stator and obtain the analytical form of single stator torque-speed relationship by fitting this normalized torque-speed curve (see also Box 1D),

$$\tau = \begin{cases} \tau_{knee} - \tau_{knee} \dfrac{(v - v_{knee})}{v_L - v_{knee}}, \text{for } v_L > v \geq v_{knee} \\ \tau_L - (\tau_L - \tau_{knee}) \dfrac{v_{knee}}{v}, \text{for } v_{knee} > v > -v_L \end{cases} \qquad [1]$$

for CCW state, and

$$\tau = \begin{cases} -\tau_L \dfrac{(v + v_L)}{v_L}, \text{for } 0 > v \geq -v_L \\ -\tau_L - (\tau_L - \tau_{knee}) \dfrac{v}{v_{knee}}, \text{for } v_L > v > 0 \end{cases} \qquad [2]$$

for CW state, where $v_{knee}$, $\tau_{knee}$, $v_L$, and $\tau_L$ take the values $160 \times 2\pi (rad/s)$, $250/4.08 (k_B T / rad)$, $300 \times 2\pi (rad/s)$, and $300/4.08 (k_B T / rad)$, respectively.

For a motor under the switching process, suppose at a given time that $n_1$ of the $N$ (we use $N$ =11 in the following simulations) RSUs that are currently in contact with the $N$ stators are in active forms, then the torque balance of the system gives:

$$n_1\tau^A(v)+(N-n_1)\tau^a(v)=\zeta v \quad [3]$$

The torque required to rotate the external load comes from $n_1$ stators pushing CW and ($N$- $n_1$) stators pushing CCW. From equations [1]-[3], one can solve the present speed $v$ of the motor and thus the instant torque generated by each stator with the experimentally measured $\tau^A(v)$ and $\tau^a(v)$.

**Numerical details**

We performed stochastic simulations to evolve the motor rotation and switch dynamics simultaneously using the standard Gillespie algorithm (21). At each simulation step a random number determines which of the following transitions takes place: rotor angular movement, one of the RSUs switches conformation, CheY-P binding on/off one of the RSUs. Another random number is generated to determine the next transition time. The angular movement of the rotor is modelled as a step movement. The step size of the motor is $\Delta\theta=2\pi/26$ with single stator (22) and $\Delta\theta=2\pi/26N$ with $N$ stators (23). At any time point t, the rotor angular position $\theta(t)$ is compared with fixed stator positions (we assume stators are equally located around the periphery of the rotor) to determine which RSUs are in contact with the stators. The motor angular position can advance a step $\Delta\theta=2\pi/26N$ (or $-\Delta\theta$ depending on the rotation direction) following a Poisson process with a rate $v/\Delta\theta$, where $v$ is the instant motor speed calculated from equation [3]. The transition rates for RSU conformation switching and CheY-P binding on/off are given by the conformational spread model discussed above.

To study the load-switching dependence, we use the same strategy as in experiments (6, 7). We gradually increase the drag coefficient of the external load attached to the motor and generate long switching traces and speed records (500s long speed records for each load line). For each switching speed record, a histogram of speed is constructed. Speed histograms typically show two peaks, one each for the CW and CCW rotation modes. Next a switching interval finding algorithm (custom written in Matlab, see reference (10) for a detailed description) is applied to all the switching speed records. The finding algorithm detects the length of all the CCW/CW intervals and calculates their mean values.

## ACKNOWLEDGEMENT

Special thanks are given to G. Oster, Howard Berg, Junhua Yuan, K. Fahrner, D. Nicolau Jr and S. Nakamura for helpful discussions. F. Bai is a research fellow of the Japan Society for the Promotion of Science (JSPS). This work has been supported by Grants-in-Aid from JSPS (to FB), and National Science Foundation Grant EF-1038636 (to JX and ZW).

## CONFLICT OF INTEREST

The authors declare that they have no conflict of interest.

Figure legends

Box 1. Schematic illustration of the BFM torque generation/switching structure and the present model components. (*A*) Schematic plot of the main structural components of the BFM. In this figure some RSUs (red) are in CW state against majority of the inactive RSUs (blue) driving the motor rotating CCW. (*B*) Side-view of a putative model of the rotor switching complex. (*C*) Free energy diagram of the conformational spread model. For simplicity the free energy difference of conformational change ($E_a$) between *ab* and *Ab* is assumed to be the same as that between *AB* and *aB*. Interactions between adjacent RSUs favour pairs with the same conformation by $E_j$ compared to any unlike pair, independent of CheY-P binding. The CheY-P binding free energy is related to the CheY-P concentration through $E_c = -\ln(c / c_{0.5})$. If $E_c >$ 0, the inactive state becomes more highly populated and hence CW bias < 0.5. Similarly $E_c < 0$ implies CW bias > 0.5. (*D*) Analytical fitting of the single stator torque-speed curves in CCW state (red) and CW state (blue). (*E*) Stator torque lowers the activation energy barrier for those contacting RSUs, and increases the basic flipping rates $\omega_{flip}$ in the conformational spread model.

Figure 1. Simulation results on the load-switching dynamics. (*A*) A typical switching angle trace of the BFM (sampled at 3000Hz) predicted by our model. The simulation is done with model parameters: $E_a = 1\ k_BT$, $E_j = 4\ k_BT$, $\omega_0 = 2100\ s^{-1}$, $c = 0.9c_{0.5}$, $\delta = 0.04$, drag coefficient of the external load $\zeta = 0.05 k_B T \times s / rad^2$, as in ref. (10). (*B*) The speed-time trace of the switching angle trace shown in (*A*). (*C-D*) Comparison of the experimental and simulated switching rate-load dependence in both CW and CCW directions. Experimental data points are taken from reference (15) with permission. The simulations are performed with increasing external load $\zeta = 0.05, 0.1, 0.3, 0.5, 1, 3, 10, 15\ k_B T \times s / rad^2$.

Figure 2. Mechanism of the load-switching dynamics. (*A*) Two time scales in the model, $t_{step}$ and $t_{flip}$ show opposite dependence on motor speed *v*. (*B*) A cartoon plot showing stator acting on more RSUs with increasing motor speed. (*C*) Average number of actual conformational changes of a randomly selected RSU on the ring in 1 second. (*D*) Snapshots of the ring activity calculated with our model (4s long, sampling rate 50000 Hz) with small ( $\zeta = 0.05 k_B T \times s / rad^2$ ), medium ($\zeta = 1 k_B T \times s / rad^2$), and big ($\zeta = 15 k_B T \times s / rad^2$) external load, respectively.

Figure S1 Model predicted stator-switching relationship. (*A*) & (*B*): $k_{CCW \to CW}$ and $k_{CW \to CCW}$ dependence on stator number simulated with 3, 5, 7, 9, 11, 13 stators in the system with small external load $\zeta = 0.1 k_B T \times s / rad^2$. (*C*) & (*D*): same as (*A*) & (*B*) but with a big external load $\zeta = 10 k_B T \times s / rad^2$.

Figure S2 Justification of the mean field approximation of the motor torque calculation. (A) Motor torque-speed curves measured with increasing stator number

(reproduced from reference (5) with permission). (B) The torque-speed curves from (A) normalized by the stator number.

Figures

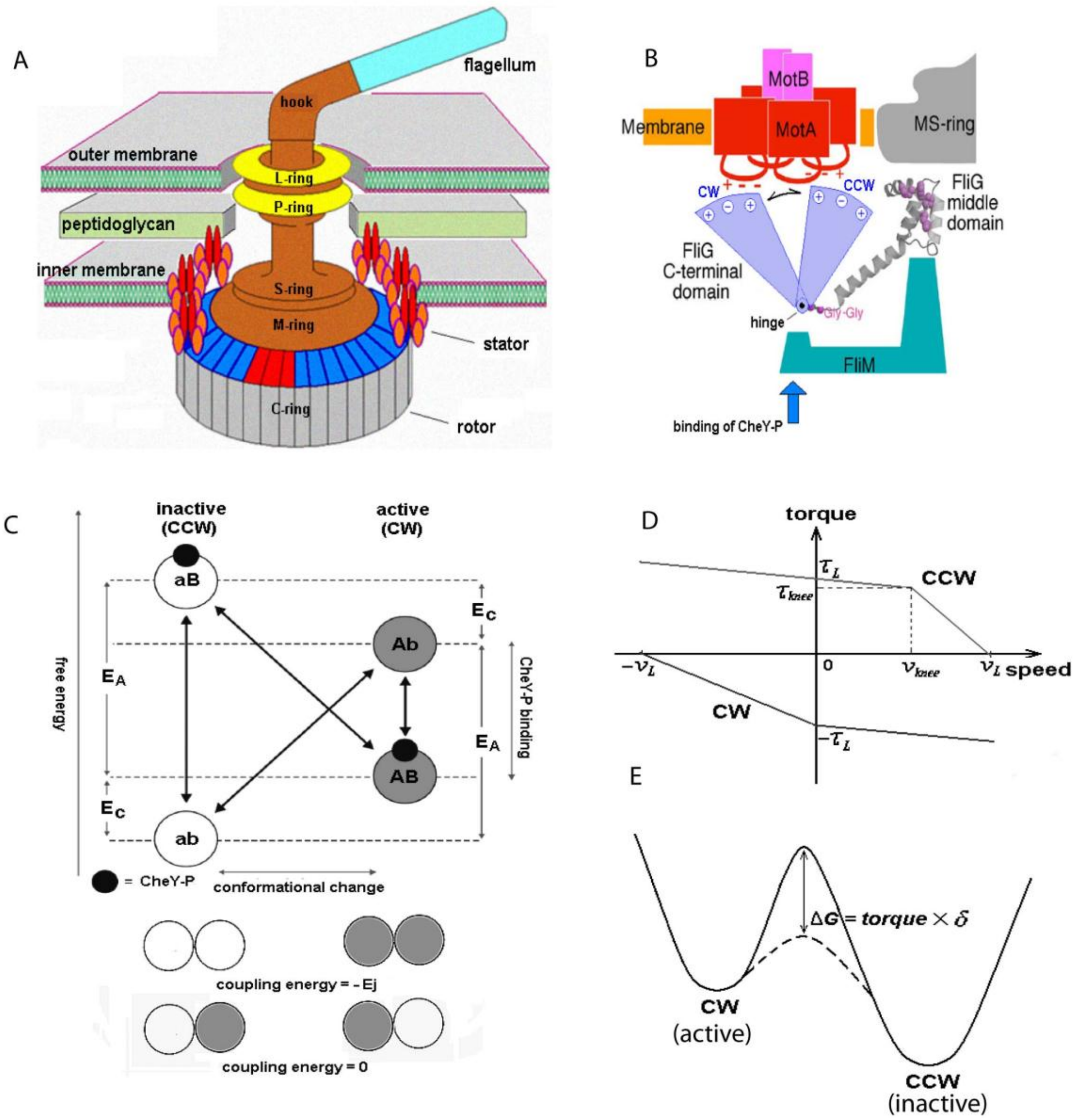

Box 1

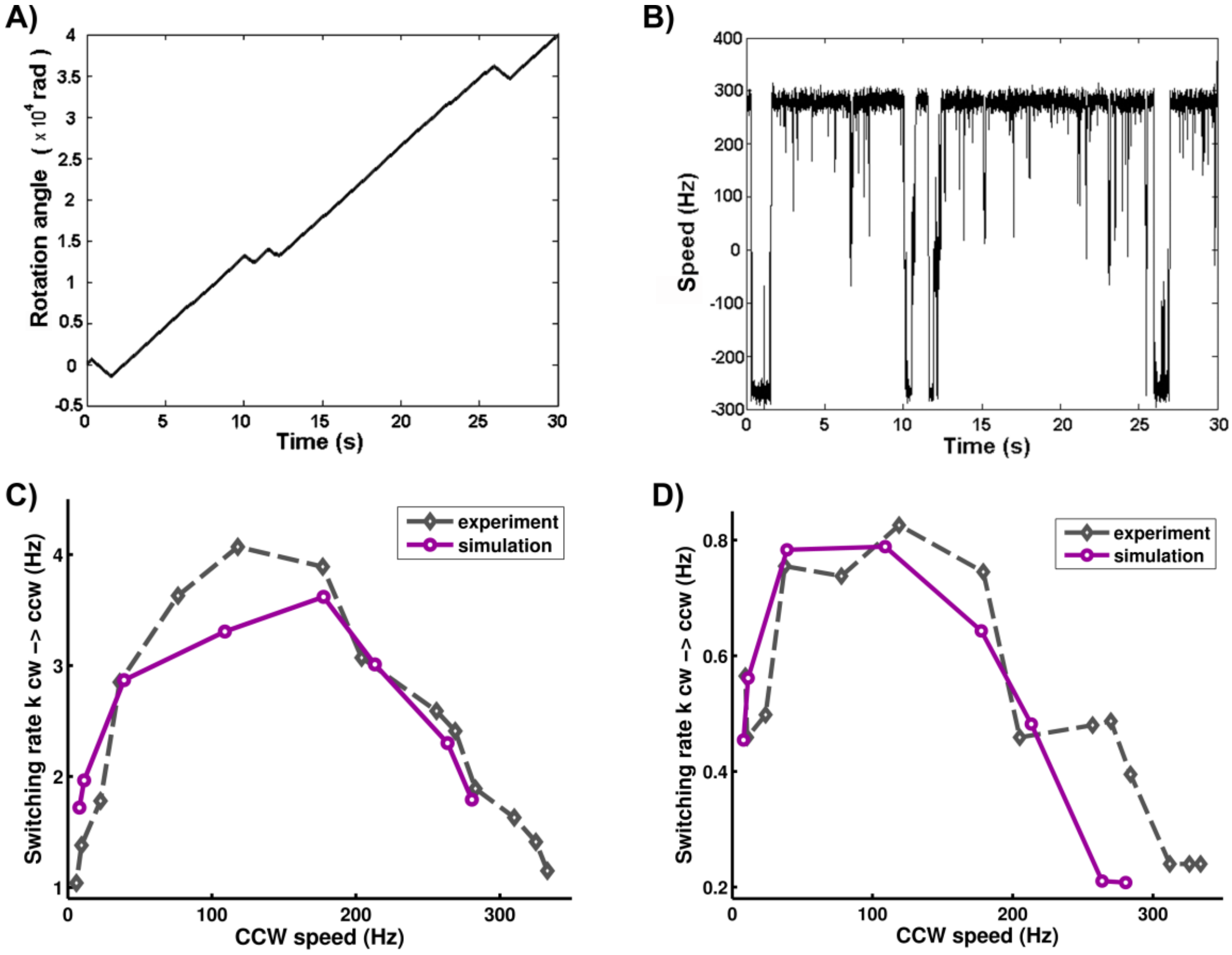

Figure 1

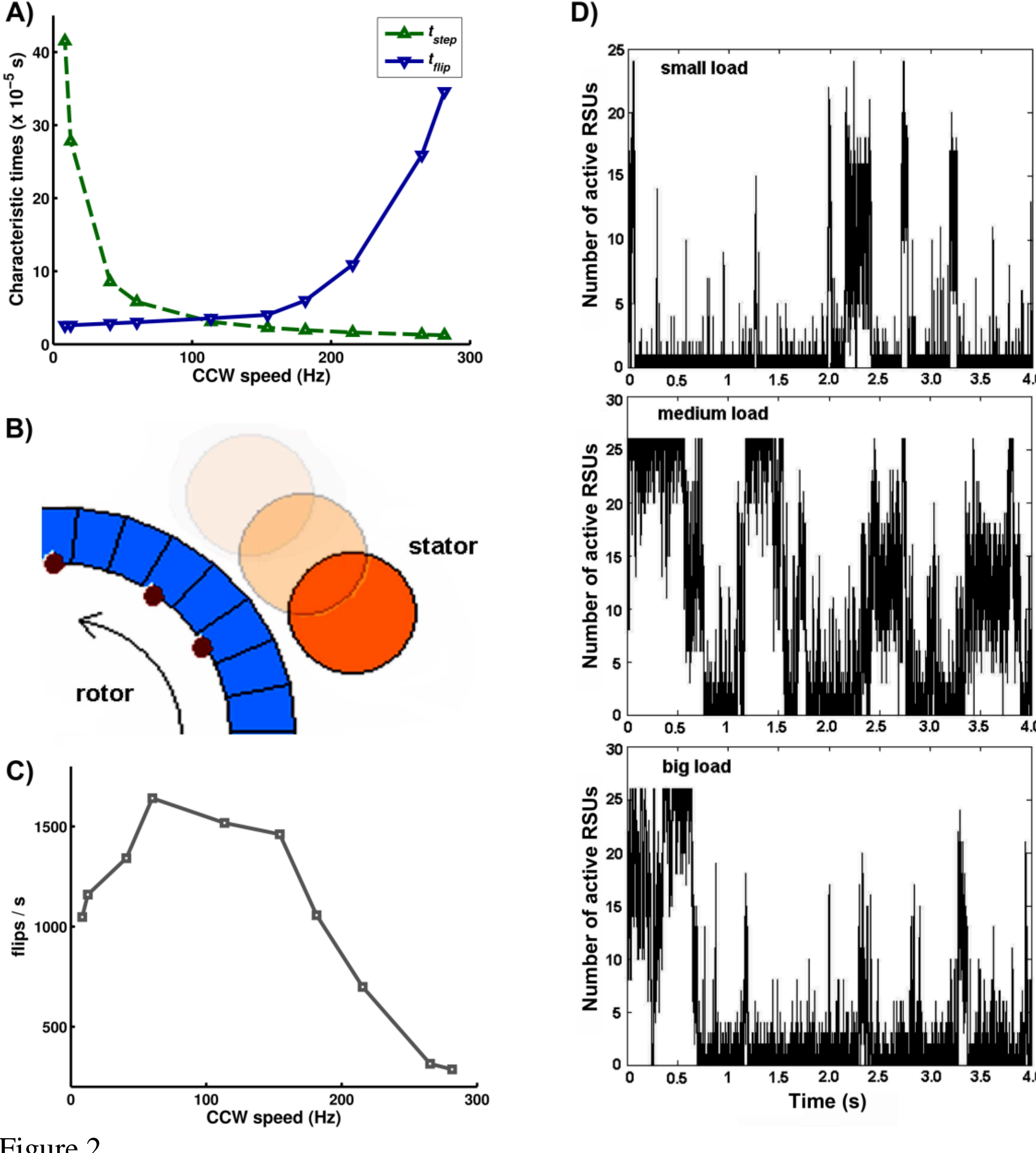

Figure 2

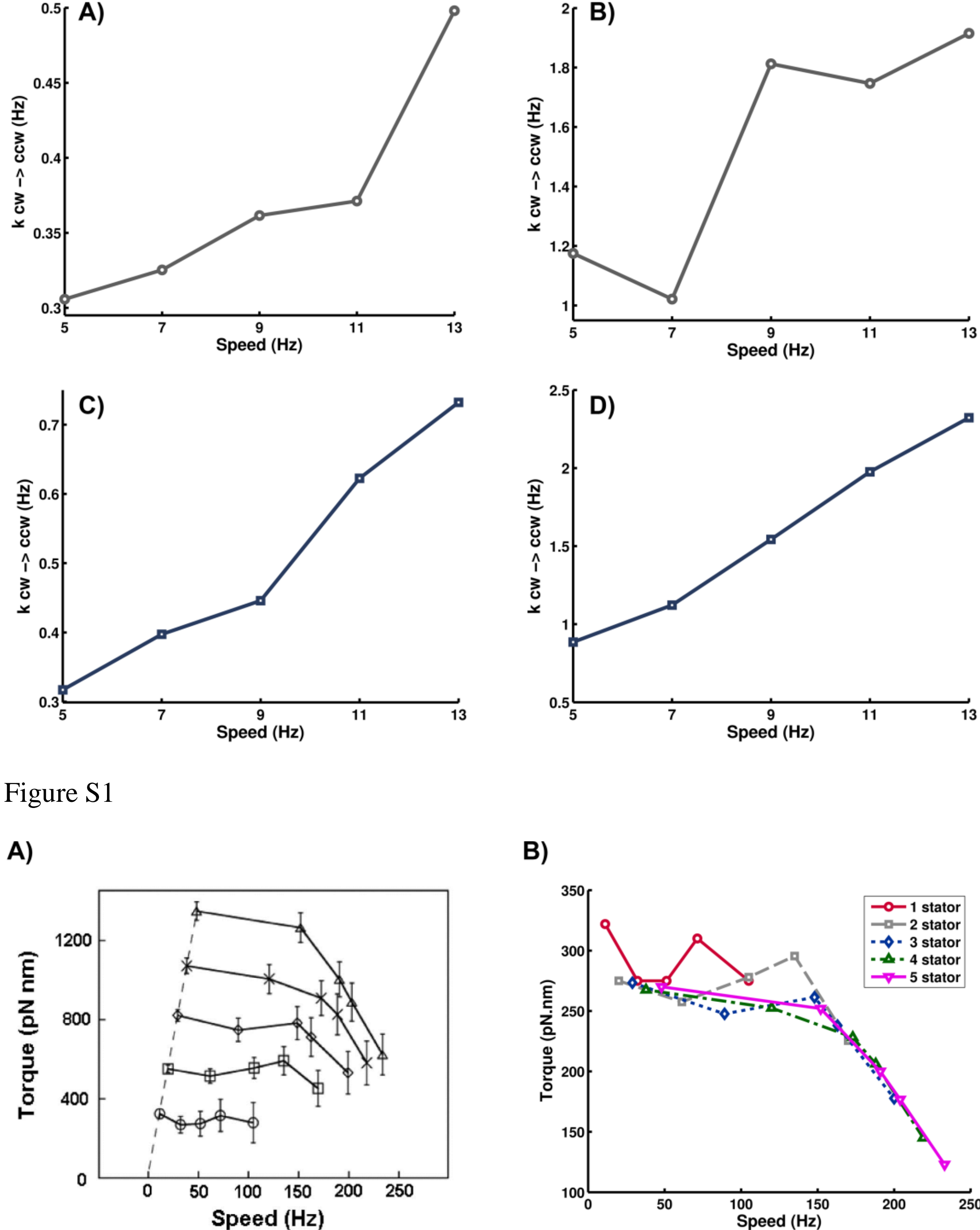

Figure S1

Figure S2